\documentclass[aps, prl, preprint, superscriptaddress, showkeys]{revtex4-1}

\usepackage{hyperref}
\usepackage{graphicx}
\usepackage{bm}
\usepackage{bbold}
\usepackage{multirow}
\usepackage{amsmath}
\usepackage{textcomp}
\usepackage{color}

\usepackage[T1]{fontenc}     
\usepackage[utf8]{inputenc} 
\usepackage{lmodern}

\begin{document}


\title{Orbitally-resolved ferromagnetism of monolayer CrI$_3$}



\author{I. V. Kashin}
\email[E-mail address: ]{I.V.Kashin@urfu.ru}
\affiliation{Theoretical Physics and Applied Mathematics Department, Ural Federal University, 620002 Ekaterinburg, Russia}

\author{V. V. Mazurenko}
\affiliation{Theoretical Physics and Applied Mathematics Department, Ural Federal University, 620002 Ekaterinburg, Russia}

\author{M. I. Katsnelson}
\affiliation{Theoretical Physics and Applied Mathematics Department, Ural Federal University, 620002 Ekaterinburg, Russia}
\affiliation{Institute for Molecules and Materials, Radboud University, Heijendaalseweg 135, 6525 AJ Nijmegen, Netherlands}

\author{A. N. Rudenko}
\affiliation{School of Physics and Technology, Wuhan University, Wuhan 430072, China}
\affiliation{Theoretical Physics and Applied Mathematics Department, Ural Federal University, 620002 Ekaterinburg, Russia}
\affiliation{Institute for Molecules and Materials, Radboud University, Heijendaalseweg 135, 6525 AJ Nijmegen, Netherlands}



\date{\today}

\begin{abstract}
Few-layer CrI$_3$ is the most known example among two-dimensional (2D) ferromagnets, which have attracted growing interest in recent years.
Despite considerable efforts and progress in understanding the properties of 2D magnets both from theory and experiment, the mechanism behind the formation of in-plane magnetic ordering in chromium halides is still under debate.
Here, we propose a microscopic orbitally-resolved description of ferromagnetism in monolayer CrI$_3$. Starting from first-principles calculations, we construct a low-energy model for the isotropic Heisenberg exchange interactions. We find that there are two competing contributions to the long-range magnetic ordering in CrI$_3$: (i) Antiferromagnetic Anderson's superexchange between half-filled $t_{2g}$ orbitals of Cr atoms; and (ii) Ferromagnetic exchange governed by the Kugel-Khomskii mechanism, involving the transitions between half-filled $t_{2g}$ and empty $e_g$ orbitals. Using numerical calculations, we estimate the exchange interactions in momentum-space, which allows us to restore the spin-wave spectrum, as well as estimate the Curie temperature. Contrary to the nearest-neighbor effective models, our calculations suggest the presence of sharp resonances in the spin-wave spectrum at 5--7 meV, depending on the vertical bias voltage. Our estimation of the Curie temperature in monolayer CrI$_3$ yields 55--65 K, which is in good agreement with experimental data. 
\end{abstract}


\maketitle

\newcommand{\ud}{\mathrm{d}}
\renewcommand{\Im}{\mathrm{Im}}
\newcommand{\TrL}{\mathrm{Tr_{L}}}
\newcommand{\TrLS}{\mathrm{Tr_{LS}}}
\newcommand{\ttwog}{$\mathrm{T_{2g}}$}
\newcommand{\tildettwog}{$\mathrm{\tilde{T}_{2g}}$}
\newcommand{\eg}{$\mathrm{E_{g}}$}
\newcommand{\kmesh}{${\bm k}$-mesh }
\renewcommand{\i}{\mathrm{i}}
\newcommand{\irrepE}{${\cal{E}}$}
\newcommand{\irrepAone}{${\cal{A}}_1$}
\newcommand{\irrepAtwo}{${\cal{A}}_2$}
\newcommand{\SetOne}{${\mathbb{1}}$~}
\newcommand{\SetTwo}{${\mathbb{2}}$~}
\newcommand{\SetThree}{${\mathbb{3}}$~}
\newcommand{\OrbOne}{$\mathrm{d_{3z^2 \textrm{-} r^2}}$~}
\newcommand{\OrbTwo}{$\mathrm{d_{x^2 \textrm{-} y^2}}$~}
\newcommand{\OrbThree}{$\mathrm{d_{xy}}$~}
\newcommand{\OrbFour}{$\mathrm{d_{xz}}$~}
\newcommand{\OrbFive}{$\mathrm{d_{yz}}$~}

\section{Introduction}

Recent discovery of ferromagnetism in two-dimensional (2D) materials has triggered enormous interest to these materials from the research community \cite{CrI3_Tc_exp, CrI3_ElF_1, CrI3_ElF_2, CrI3_ElF_3, NatureElectronics, Nature_01}. Apart from being fundamentally interesting object in the context of the Mermin-Wagner theorem, which forbids magnetic ordering in 2D at any nonzero temperature for the case of isotropic Heisenberg or easy-plane-anisotropic magnets \cite{PhysRevLett.17.1133}, 2D magnets are especially promising for applications, such as spintronics, quantum computing and energy-efficient electronics \cite{Burch2018,Gong2019,Cortie2019,Gibertini2019}. 

A few-layer chromium triiodide (CrI$_3$) is the most studied representative among 2D ferromagnets with the Curie temperature in the range 45--61 K, depending on the layer thickness \cite{CrI3_Tc_exp}. It was first synthesized by mechanical exfoliation of bulk CrI$_3$ crystal, and identified as an Ising ferromagnet with out-of-plane spin orientation \cite{CrI3_Tc_exp}. Being strongly sensitive to external electric field, few-layer CrI$_3$ became the subject of numerous experimental studies \cite{CrI3_Tc_exp, Paper_02, Experiment_01, Klein1218, NatComm.9.2516.2018}, particularly aimed to electrical control and manipulation of magnetic states \cite{CrI3_ElF_1, CrI3_ElF_2, CrI3_ElF_3}. 
It was found that the magnetism of few-layer CrI$_3$ is essentially layer-dependent: monolayer and trilayer CrI$_3$ demonstrate ferromagnetic (FM) ordering in its ground state, whereas bilayer turns out to be antiferromagnetic (AFM) \cite{CrI3_ElF_1, CrI3_ElF_2, Klein1218, NatComm.9.2516.2018}. This behavior opens up the possibility
to realize an electrical switching of magnetic ordering in bilayer CrI$_3$ \cite{CrI3_ElF_1,CrI3_ElF_3}.
As it has been shown in Ref. \cite{Paper_02}, the stacking order and interlayer exchange interactions in CrI$_3$ are intimately connected,
leading to stacking-dependent magneto-Raman signatures \cite{Paper_10}. 
Despite the similarities between monolayer and trilayer CrI$_3$, the latter appears to be closer to the bulk crystal \cite{CrI3_Tc_exp}. This makes ultra-thin films with thickness of one or two atoms the most interesting objects for further studies.

On the theory side, a large number of studies was devoted to a microscopic analysis of magnetism in 2D CrI$_3$ and its analogues \cite{Paper_01, Paper_04, Paper_05, Paper_06, Paper_07, Theory_01, Theory_02, Theory_03_olsen}. 
Most of them were based on first-principles calculations (density functional theory and beyond) with direct mapping on the Heisenberg model. Being a standard approach in material science, such method allows one to describe the macroscopic quantities (e.g., the Curie temperature) with reasonable accuracy \cite{Lado_2017, PhysRevLett.121.067701, Torelli_2018}. On the other hand, this method is limited with respect to the orbital-resolved physics, and does not provide insights into the origin of exchange interactions.
In Ref. \cite{Paper_01}, an important role of the Coulomb interactions in 2D magnets was revealed using the combination of first-principles calculations with classical Monte Carlo simulations. An essential contribution of the ligand states in the formation of ferromagnetism in CrI$_3$ was emphasized in Ref.~\cite{Solovyev_CrI3}. Particularly, the authors showed that the Goodenough-Kanamori-Anderson rule for 90\textdegree~Cr-I-Cr bond cannot serve as an explanation for short-range ferromagnetic order due to the competing character of exchange interactions. Being also applicable to the monolayer case, this finding suggests an orbitally-resolved analysis to be necessary to understand the mechanism behind the magnetic ordering in CrI$_3$. An attempt to provide an orbitally-resolved picture was first made in Ref.~\cite{Jij_Bilayer} in the context of interlayer interactions in bilayer CrI$_3$. 
However, a systematic orbital theory of the \emph{intralayer} ferromagnetic ordering in CrI$_3$ was not formulated up to now. 

\begin{figure}[h!]
\centering
\includegraphics[width=0.95\textwidth,angle=0]{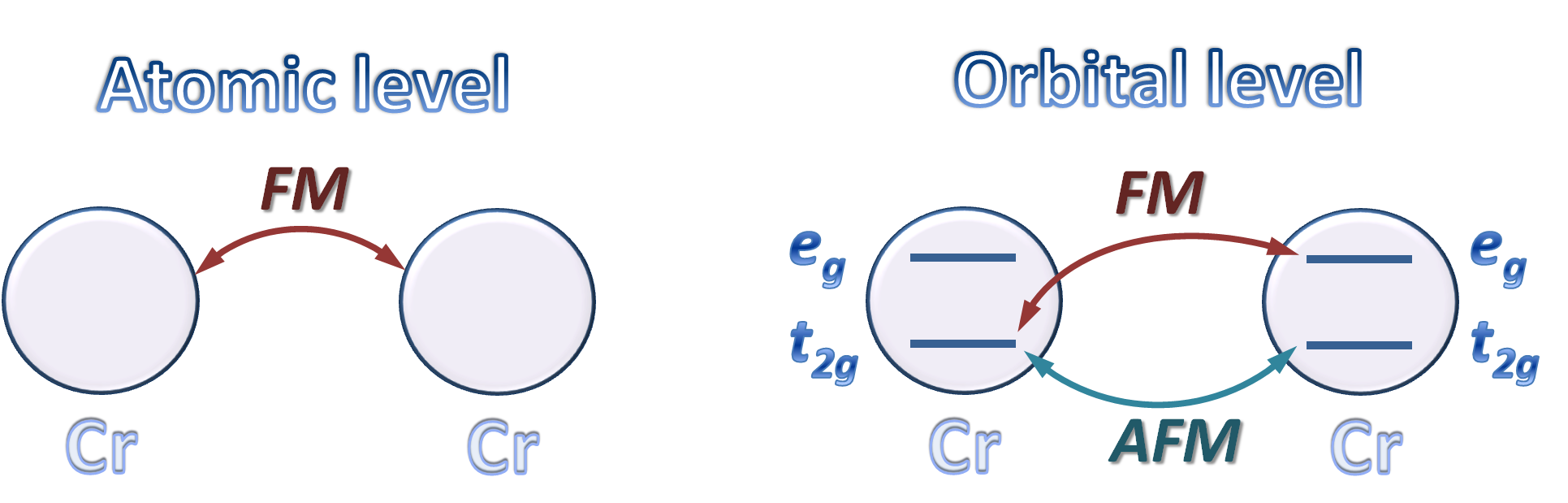}
\caption{Schematic illustration of the main idea of our study. Contrary to the atomic level (left panel) employed in previous studies, here we work at the level of individual orbital states of chromium atoms (right panel). This allows us to reveal a rich picture of competing ferromagnetic ($t_{2g}$ - $e_{g}$) and antiferromagnetic ($t_{2g}$ - $t_{2g}$) interactions in monolayer CrI$_3$.}
\label{Idea}
\end{figure}
 
In this paper, we propose a microscopic, orbitally-resolved picture of ferromagnetism in CrI$_3$. Starting from first-principles electronic structure calculations, we construct a microscopic model allowing us to determine orbitally-resolved exchange interactions by utilizing the magnetic force theorem. 
We find two competing contributions to the magnetic interactions in CrI$_3$ originating from the AFM and FM coupling between symmetry-equivalent ($t_{2g}$--$t_{2g}$) and inequivalent ($t_{2g}$--$e_g$) orbitals, respectively (see Figure~\ref{Idea} for a schematic illustration). Although the first contribution, which is the well-known Anderson's superexchange, is typical to many magnetic materials, the second contribution is less common and can be understood in terms of the Kugel-Khomskii picture. In this picture, the ferromagnetic interaction arises from 
indirect transitions between the half-filled ($t_{2g}$) and unoccupied ($e_g$) orbitals modulated by the intraatomic Hund's rule coupling. We also qualitatively determine the role of external electric field and substrate screening in the formation of magnetic ordering in monolayer CrI$_3$.

\section{Method}

\newcommand{\Umatrix}{U}

Our computational approach is based on first-principles electronic structure calculations within the density functional theory (DFT) \cite{dft}. We start from the projected augmented wave (PAW) method \cite{paw1,paw2} as implemented in \emph{Vienna ab-initio simulation package} ({\sc vasp}) \cite{vasp1,vasp2}.
To describe the exchange-correlation effects, we use the generalized gradient approximation (GGA) \cite{pbe} augmented by the Hubbard-$U$ corrections (GGA+$U$ method \cite{Anisimov,Dudarev}), where $U$ is treated as a free parameter.
An energy cutoff of 250 eV was used for the plane wave basis. The convergence threshold during the self-consistent solution of the Kohn-Sham equations was set to 10$^{-10}$ eV. A (8$\times$8) {\bf k}-point mesh was used to sample the Brillouin zone. To avoid spurious interactions 
between the cells, a vacuum slab of 20~\AA~was added in the direction normal to the 2D plane. The atomic structure and lattice parameters were fully relaxed preserving the lattice symmetry ($D_{3d}$ point group) until the residual forces were less than 0.001 eV/\AA. The resulting lattice parameter is found to be $a=4.03$ \AA, while the layer thickness (separation between topmost and lowermost iodine atoms) is $d=3.03$ \AA.

As a next step, we construct an extended tight-binding model by mapping the resulting DFT wave functions onto the Wannier functions \cite{Marzari2012}. To this end, we use the {\sc wannier90} package \cite{wannier902}. The tight-binding model was constructed in the canonical basis of cubic harmonics, corresponding to chromium $3d$, and iodine $5s$ and $5p$ orbitals. 
The tight-binding Hamiltonian considered in our study has the following form:
\begin{eqnarray}
\label{Ham_with_VertBias}
\hat{H} = \sum_{\alpha \beta, \, \sigma}
t^{\sigma}_{\alpha \beta} \, \hat{c}^{\dagger}_{\alpha \sigma} \hat{c}_{\beta \sigma}
+
\frac{V}{d} \, \sum_{\alpha, \, \sigma}
\hat{z}_{\alpha} \, \hat{n}_{\alpha \sigma} ,
\end{eqnarray}
where 
$\hat{c}^{\dagger}_{\alpha \sigma}$ ($\hat{c}_{\beta \sigma}$) is is the creation (annihilation) operator of an electron with spin $\sigma$ at orbital $\alpha$ ($\beta$), and
$t^{\sigma}_{\alpha \beta}$ is the spin-resolved matrix of hopping integrals. The second term in Eq.~(\ref{Ham_with_VertBias}) describes the effect of external static electric field \cite{PhysRevB.98.201401,PhysRevB.99.064513}, where
$\hat{z}_{\alpha}$ is the $z$-component of the position operator associated with orbital $\alpha$,
$\hat{n}_{\alpha \sigma}=\hat{c}^{\dagger}_{\alpha\sigma}\hat{c}_{\alpha\sigma}$ is the particle number operator,
$V$ is the vertical bias potential, and $d$ is an effective 2D layer thickness. It should be noted that Eq.~(\ref{Ham_with_VertBias}) describes the electric field effect only qualitatively. In our study, we consider two cases only: $V=0$ and $V=1$ eV. The latter corresponds to an unscreened vertical electric field of $E_z=V/d\approx$ 0.33 eV/\AA.


In order to calculate the magnetic interactions in CrI$_3$, we use the Green's functions formalism allowing us to apply the magnetic-force theorem \cite{InfsSpRot}. Within this scheme, the Heisenberg exchange interactions are expressed via infinitesimal spin rotations $J_{ij}=\partial^2 E / \partial \phi_i \partial \phi_j$ as:
\begin{eqnarray}
\label{Exch_InfSpRot}
J_{ij} = \frac{1}{8 \pi} \, \Im \int_{-\infty}^{E_{F}} \TrL \, \Big[ \sum_{\sigma} \Delta_{i} \, G^{\sigma}_{ij} \, \Delta_{j} \, G^{-\sigma}_{ji} \Big] \, \ud \omega,
\end{eqnarray}
where 
$G^{\sigma}_{ij} = \sum_{\bm{k}} (\omega - H_{ij}^{\sigma})^{-1} \cdot e^{i \bm{k} {\bf T}_{ij}}$ is the spin-polarized intersite Green's function, taken as an integral over the Brillouin zone, with $H_{ij}^{\sigma}$ being the Hamiltonian matrix (Eq.~ (\ref{Ham_with_VertBias})), ${\bf T}_{ij}$ is the translation vector, connecting atoms $i$ and $j$, and $\bm{k}$ is the reciprocal lattice vector.
$\Delta_{i} = H_{ii}^{\uparrow} - H_{ii}^{\downarrow}$ is the intraatomic exchange splitting  of the Cr$_{i}$($3d$) shell,
$\TrL$ denotes the trace over the orbital ($L$) indices, and
$E_{F}$ is the Fermi energy.  
Hereinafter positive/negative $J_{ij}$ corresponds to FM/AFM ordering, respectively.

The exchange interactions defined above represent a reliable starting point for the analysis within the isotropic Heisenberg model:
$H=-\sum_{ij}J_{ij}{\bf S}_i{\bf S}_j$. However, the advantage of the approach presented is that it allows one to decompose $J_{ij}$ into the individual orbital contributions. 
To perform this decomposition, we define the exchange splitting matrix in the orbital space, and perform its diagonalization, $\Delta_{i}^{m \, m'}=U_{i}^{m \, k} \cdot \widetilde{\Delta}_{i}^{k \, k} \cdot (U_{i}^{k \, m'})^{*}$,
where $\widetilde{\Delta}_{i}^{kk}$ is the on-site matrix, which is diagonal in the orbital space and 
the matrix $U$ is a unitary transformation matrix which defines the basis where $\Delta$ is diagonal.
The exchange interaction between the {\it k}-th orbital at site $i$ and the {\it k$^{\prime}$}-th orbital at 
site $j$ is then given by \cite{Mazurenko_Na2V3O7, PhysRevB.89.214422, PhysRevLett.116.217202}
\begin{eqnarray}
J^{kk'}_{ij} = \frac{1}{8 \pi} \, \Im \int_{-\infty}^{E_{F}} \TrL \, \Big[ \sum_{\sigma} \widetilde{\Delta}^{kk}_{i} \cdot
\widetilde{G}^{kk'}_{ij \, \sigma} \cdot \widetilde{\Delta}^{k'k'}_{j} \cdot \widetilde{G}^{k'k}_{ji \, -\sigma} \Big] \, \ud \omega,
\label{Eq3}
\end{eqnarray}
where $\widetilde{G}^{kk'}_{ij \, \sigma}= \sum_{m m'} U_{i}^{k \, m} \cdot 
G^{mm'}_{ij \, \sigma} \cdot (U_{j}^{m' \, k'})^{*}$. 
To perform an accurate evaluation of the exchange interactions by means of Eqs.~(\ref{Exch_InfSpRot}) and (\ref{Eq3}), we use a ($400 \times 400$) {\bf k}-point mesh for calculating the Brillouin zone integrals,
which was checked to be sufficient to reach numerical convergence.

The Curie temperature is estimated in our study at the level of the random-phase approximation (RPA), which is also known as the Tyablikov's approximation \cite{Tyablikov}, and spherical model \cite{White}. This approach provides a reasonable description of the long-wavelength spin fluctuations, which is the crucial factor suppressing the ordering temperature in 2D \cite{Irkhin}. In case of multiple magnetic sublattices, the critical temperature of magnetic ordering can be determined from the following system of equations \cite{PhysRevB.71.174408,PhysRevB.78.104402}:
\begin{eqnarray}
\label{Formula_Tc_RPA}
T_{C} = \frac{2}{3}  \frac{S_{a} (S_{a} + 1)}{\langle s_{a}^{z} \rangle}
\left\{ \frac{1}{\Omega} \int \mathrm{d} {\bf{q}} \, [N^{-1}({\bf{q}})]_{aa} \right\} ^{-1} ,
\end{eqnarray}
where ${\bf{q}}$ is the magnon wave vector, $S_{a}$ is the total spin at sublattice $a$, 
$\langle s_{a}^{z} \rangle$ is the averaged sublattice magnetization projected on the $z$-axis for sublattice $a$, $\Omega$ is the volume of the Brillouin zone, and $N_{ab}({\bf{q}})$ is a matrix given by:
\begin{eqnarray}
\label{Formula_Nq}
N_{ab}({\bf{q}}) = \delta_{ab} \Big( \Lambda + \sum_{p} {\tilde{J}}_{ap}({\bf{0}}) \langle s_{p}^{z} \rangle \Big) - \langle s_{a}^{z} \rangle {\tilde{J}}_{ab}({\bf{q}}),
\end{eqnarray}
where $\Lambda$ is the magnetic anisotropy energy (MAE) and ${\tilde{J}}_{ab}({\bf{q}})$ is the Fourier component of the normalized exchange parameter ${\tilde{J}}_{ij} = J_{ij} / (S_{i} S_{j})$. The spectrum of spin-wave excitations $\epsilon({\bf q})$ can be obtained by solving the secular equation $\mathrm{det}[N({\bf{q}})-\epsilon({\bf q})\mathcal{I}]=0$ with $\mathcal{I}$ being the unity matrix. In the case of a hexagonal magnetic lattice with nearest-neighbor exchange interactions, this approach leads to a dependence in the form $T_{C} \sim JS^2/\mathrm{ln}(\frac{JS^2}{\Lambda})$ \cite{PhysRevApplied.11.064015}. We note that a more accurate estimate of the Curie temperature for 2D magnets can be done within the self-consistent spin-wave theory with fluctuation corrections \cite{Irkhin}. Within the accuracy of the main logarithm [$\mathrm{ln}(\frac{J}{\Lambda})$] and for spins $S$ > 1/2, the results are close to the approach used here.


\section{Results and discussion}

\begin{figure}[h!]
\centering
\includegraphics[width=0.95\textwidth,angle=0]{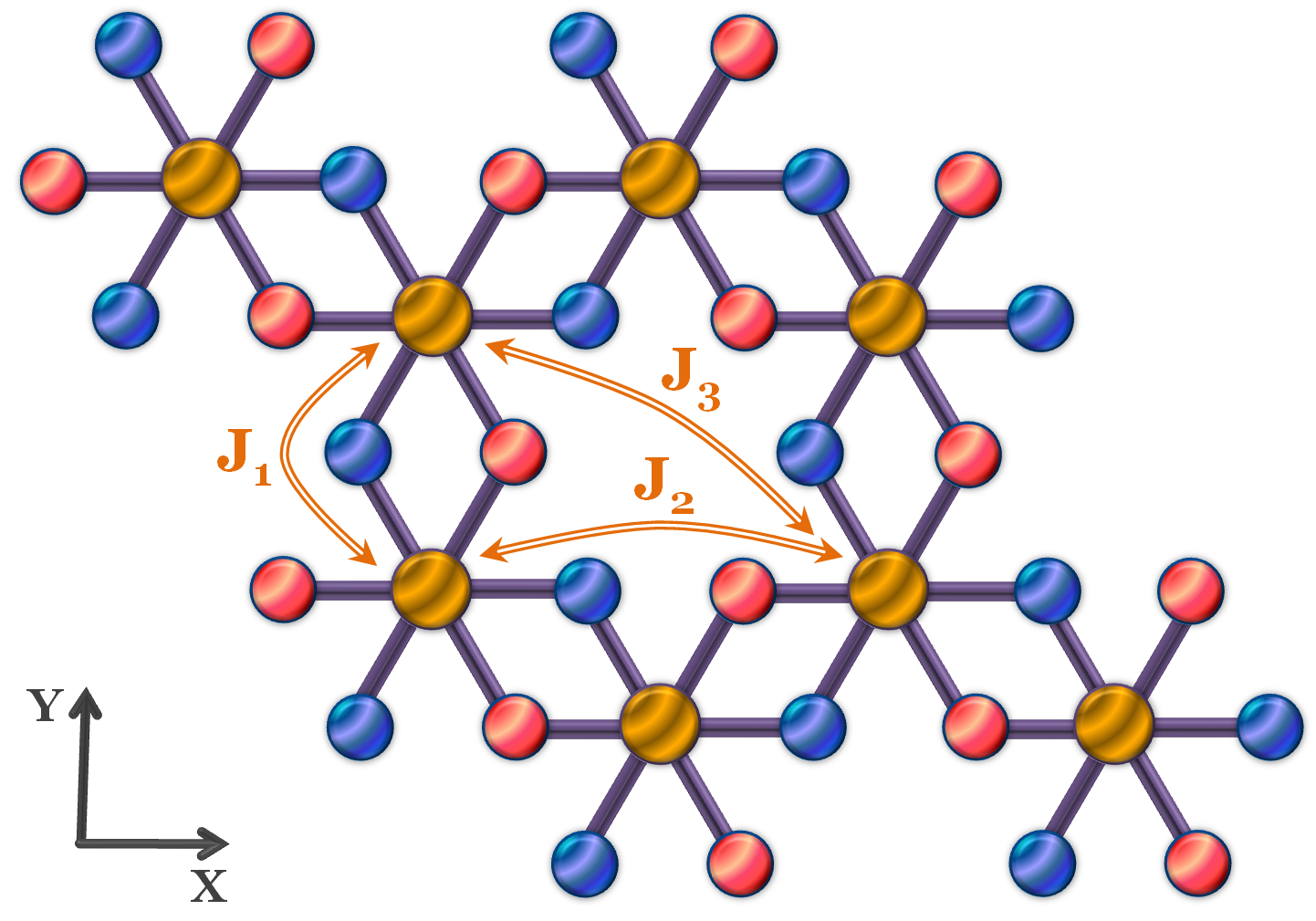}
\caption{Schematic crystal structure of CrI$_3$ monolayer. Orange circles correspond to chromium atoms. Blue and red circles denote the iodine atoms higher and lower of Cr sublattice, respectively. Vertical displacement between iodine sublattices is 3.03 \AA.  Arrows denote first-, second- and third-nearest-neighbor exchange interactions.}
\label{CrI3_Crystal_Structure}
\end{figure}

\begin{figure}[h!]
\centering
\includegraphics[width=0.95\textwidth,angle=0]{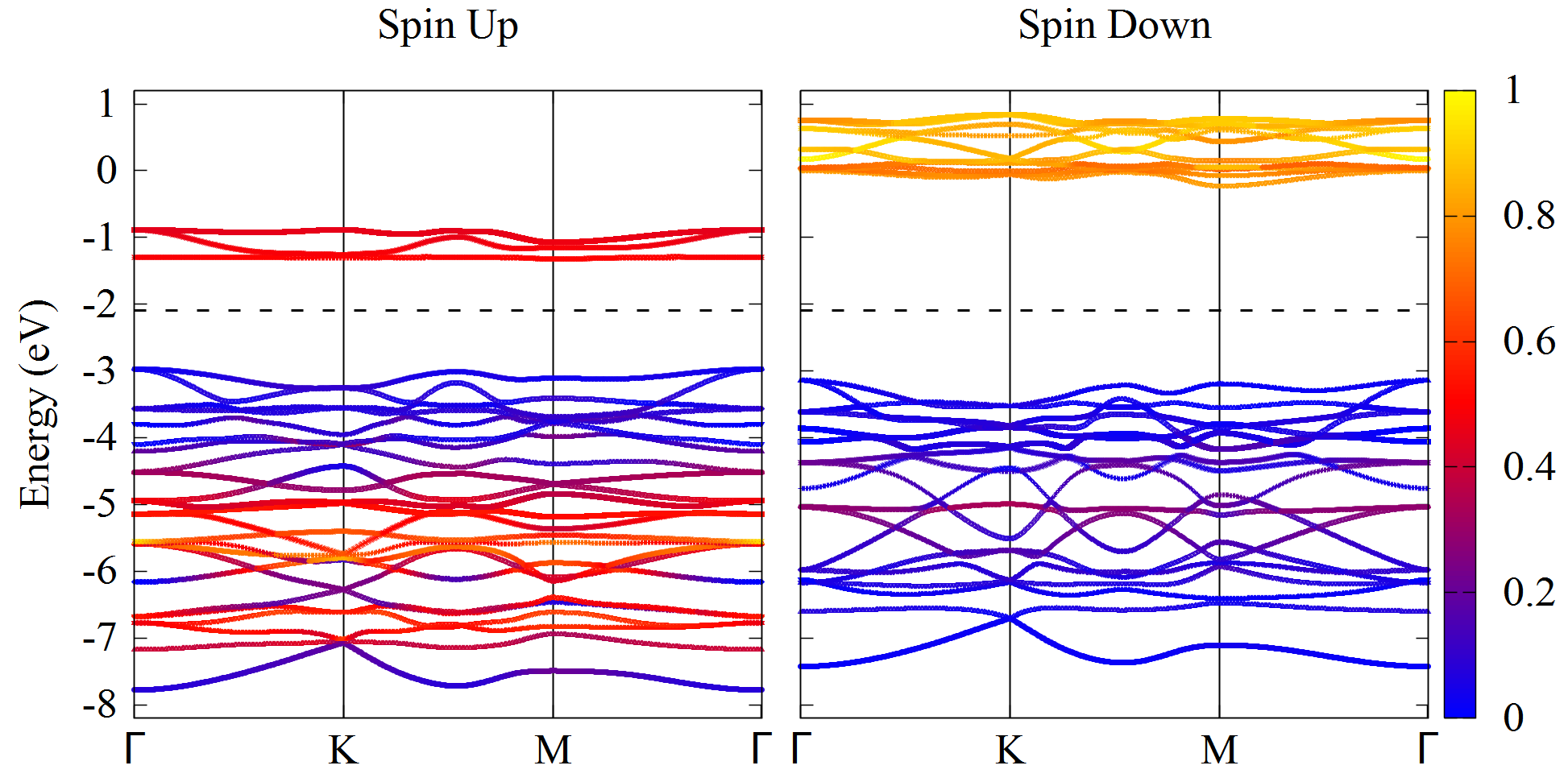}
\caption{Band structure of CrI$_3$ monolayer, obtained from GGA+\textit{U} (\textit{U} = 3 eV) calculations. The relative contribution of Cr $3d$ states is illustrated by color.}
\label{Bands_Comparing}
\end{figure}

\begin{figure}[h!]
\centering
\includegraphics[width=0.95\textwidth,angle=0]{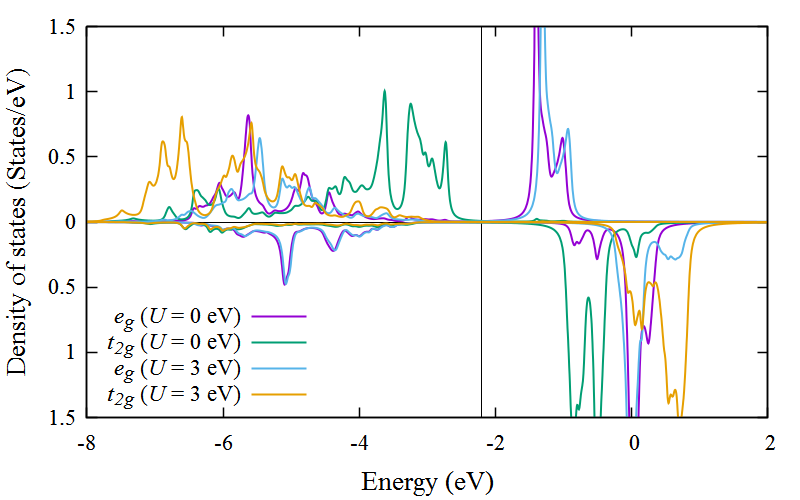}
\caption{Partial density of states of Cr $3d$ shell obtained from low-energy models without local Coulomb interaction at Cr atoms and with \textit{U} = 3 eV.}
\label{DoS_plot}
\end{figure}

Monolayer CrI$_3$ is a hexagonal crystal with two sublattices forming a honeycomb structure of Cr atoms. Each Cr atom is located in the octahedral environment of six I atoms (see Figure \ref{CrI3_Crystal_Structure}). In order to investigate its electronic and magnetic properties, we start from first-principles calculations within the GGA+$U$ approach. In Figure \ref{Bands_Comparing}, we show the spin-resolved band structure with relative contribution of Cr $3d$ shell, indicated by color. Apart from different values of a band gap for majority and minority spin channels, it is seen that the distribution of Cr $3d$ states is essentially different for spin up and spin down channels. While minority Cr $d$ states are empty and well separated from the $p$ states of iodine, the majority $d$ states are partially occupied and hybridized strongly with the iodine $p$ states. This suggest a nontrivial role of the ligand states in the formation of magnetic properties of CrI$_3$.
Figure \ref{DoS_plot} shows the density of electronic states projected onto different orbitals. According to the crystal symmetry, the $d$-shell of Cr atoms splits into two inequivalent sets of orbitals: three $t_{2g}$ and two $e_g$. From Fig.~\ref{DoS_plot} one can see that the $e_g$ orbitals are either filled or empty, whereas $t_{2g}$ orbitals are half-filled, making monolayer CrI$_3$ almost an ideal realization of a system with spin $S=3/2$. The splitting between the majority and minority $t_{2g}$ orbitals is governed by the Coulomb interaction $U$, whereas the position of $e_g$ states are almost unaffected by this parameter.

As it immediately follows from the band structure, the total magnetization in CrI$_3$ is 3 $\mu_B$ per Cr atom.
In order to distinguish between the contributions of $t_{2g}$ and $e_{g}$ orbitals to the magnetic moment, we consider the difference between the occupation matrices for spin up and spin down electrons \cite{PhysRevB.88.081405}, 
\begin{eqnarray}
M_{k} = -\frac{1}{\pi} \, \Im \int_{-\infty}^{E_{F}} [G_{k}^{\uparrow}(\omega) - G_{k}^{\downarrow}(\omega)] \, \ud \omega ,
\end{eqnarray}
where $G^{\sigma}_k \equiv [G^{\sigma}_{kk}]_{ii}$ is the local Green's function for orbital $k$. The resulting values are given in Table~ \ref{Occup_MagMom_T2g_Eg}), from which one can identify $t_{2g}$ as the orbitals with the largest spin polarization. A slight deviation of the total magnetic moment (3.12 $\mu_{B}$) from the expected value can be attributed to the presence of a non-negligible Cr($3d$)$-$I($5s5p$) hybridization, resulting in a compensating iodine magnetization with $-0.04~\mu_{B}$ per atom. We note that the magnetic moment appears to be almost independent of the Coulomb parameter \textit{U}. Although the spin density undergoes a redistribution in the presence of $U$, the occupation matrices are weakly affected.

\begin{table}[h!]
\centering
\caption [Bset]{Orbital occupation and magnetic moments of $t_{2g}$ and $e_{g}$ orbitals in CrI$_3$. The values are obtained from the integration of the density of states (Fig.~\ref{DoS_plot}) over occupied states.}
\label {Occup_MagMom_T2g_Eg}
\setlength{\tabcolsep}{7pt}
\begin {tabular}{c|ccc}
 \hline  \hline
 Orbital & Spin Up & Spin Down & Magnetic moment ($\mu_{B}$) \\
 \hline 
 $t_{2g}$  &   2.984   &  0.265    &   2.719  \\
 $e_{g}$   &   1.042   &  0.642    &   0.400  \\
 \hline  \hline
\end {tabular}
\end {table}

\begin{figure}[h!]
\centering
\includegraphics[width=0.95\textwidth,angle=0]{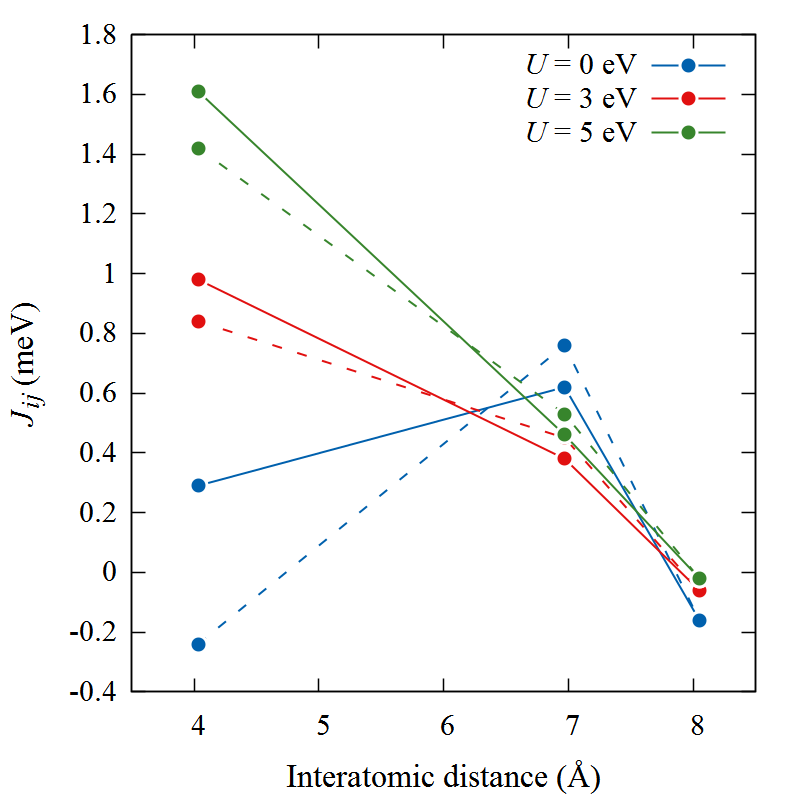}
\caption{The isotropic exchange interactions calculated between the magnetic moments of chromium atoms in CrI$_3$ as a function of the distance. Solid (dashed) lines are the results obtained in the absence (presence) of a vertical bias potential ($V=1$ eV).}
\label{Fig_JijTotalCS}
\end{figure}

The calculated interatomic exchange interactions in CrI$_3$ monolayer are presented in Fig.~\ref{Fig_JijTotalCS}. One can see a significant influence of the Coulomb interaction on the nearest-neighbor exchange interaction $J_{ij}$ between Cr atoms. One can see a gradual increase of the interaction with $U$, which can be attributed to the localization of magnetic moments. This tendency is in line with the results of hybrid functional calculations, predicting larger $J_{ij}$ values compared to standard semilocal functional.  However, we note that the obtained $J_{ij}(U)$ dependence is different from the dependence expected within the conventional superexchange mechanism, where $J_{ij}\sim 1/U$. The typical $U$ values in chromium-based compounds including CrI$_3$ multilayer is around 3 eV, as follows from the estimations within the random-phase approximation \cite{Jij_Bilayer} as well as within the linear response approach \cite{C5CP04835D}. In this case for the nearest-neighbor exchange interaction we have $J\approx1$ meV, which is reasonable agreement with the results obtained by means of DFT total energy calculations in Ref.~\onlinecite{C5TC02840J} (note different definition for the Heisenberg Hamiltonian used in this paper).

\begin{figure}[h!]
\centering
\includegraphics[width=0.95\textwidth,angle=0]{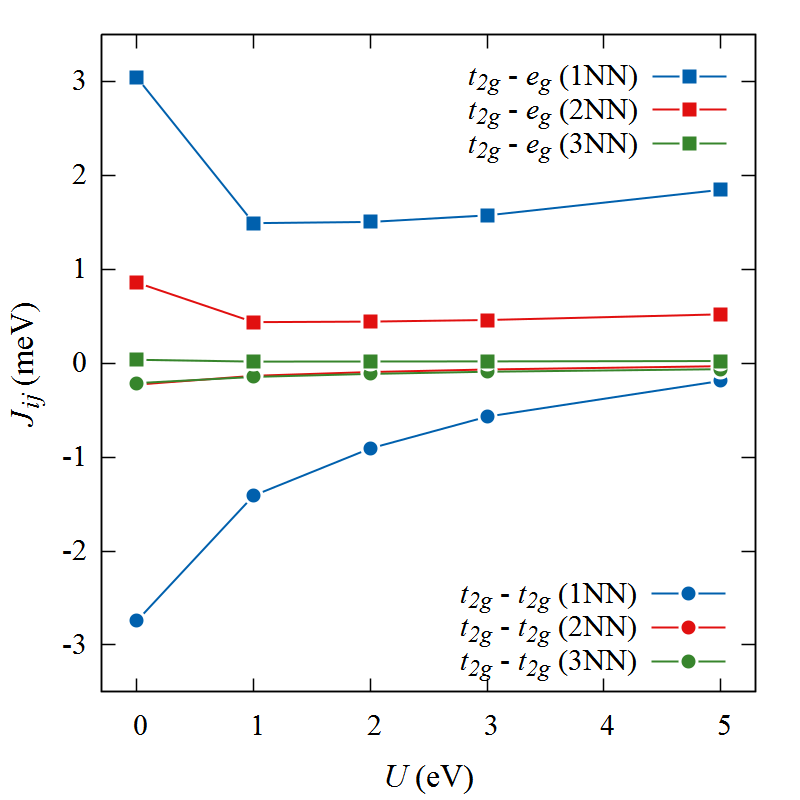}
\caption{Orbital decomposition of the total isotropic exchange interactions $J_{ij}$ in CrI$_3$ shown as a function of the Coulomb interaction $U$ for three nearest-neighbor (NN) Cr atoms. Circles and squares denote $t_{2g}-t_{2g}$ and $t_{2g}-e_{g}$ channels, respectively.}
\label{Tab_OrbDecomp_Total_per_CS}
\end{figure}

\begin{figure}[h!]
\centering
\includegraphics[width=0.95\textwidth,angle=0]{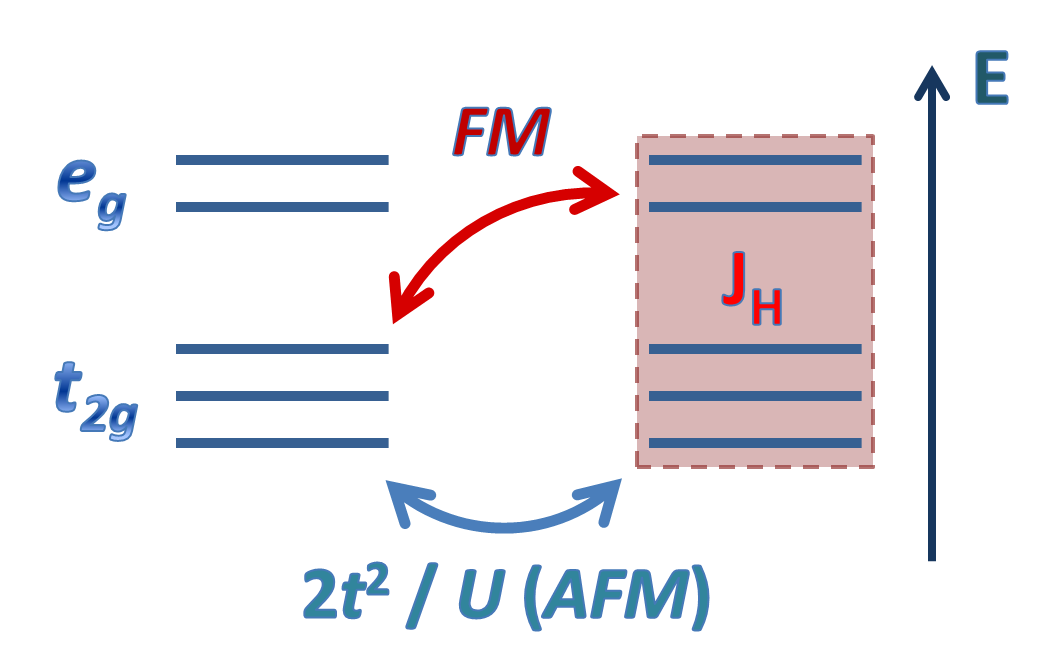}
\caption{Schematic representation of the Kugel-Khomskii interaction mechanism as applied to CrI$_3$ monolayer. FM and AFM Cr-Cr interactions are assumed to be mediated by the iodine $sp$ states.}
\label{Fig_KugelKhomskii}
\end{figure}

In order to study the driving force of CrI$_3$ ferromagnetism, we performed a decomposition of $J_{ij}$ into different orbital contributions. The results are presented in Table \ref{Jij_OrbDecomp_Table} and in Fig.~\ref{Tab_OrbDecomp_Total_per_CS}. The dominant contribution to the exchange interactions in CrI$_3$ originates from the $t_{2g}-t_{2g}$ and $t_{2g}-e_{g}$ orbital channels, demonstrating an essentially different character.
Particularly, the AFM coupling between the occupied $t_{2g}$ orbitals competes with the FM coupling between the occupied $t_{2g}$ and unoccupied $e_g$ states. This picture appears consistent with the well-known Kugel-Khomskii formalism developed on the basis of the superexchange theory \cite{KugelKhomskii, KhomskiiBook, Mazurenko_Na2V3O7}. According to this approach the total exchange interaction of a multiorbital electronic system can be presented as a sum of two contributions in the following form:  
\begin{eqnarray}
\label{KugelKhomskii_HundImpact}
J_{1} = J_{1}^{t_{2g} - t_{2g}} + J_{1}^{t_{2g} - e_{g}} =  -\frac{2(t^{t_{2g} - t_{2g}}_{1})^2}{U} + \frac{2(t^{t_{2g} - e_{g}}_{1})^2 {\cal{J}}_{H} }{(U + \Delta)(U + \Delta - {\cal{J}}_{H})}.
\label{KKh}
\end{eqnarray}
Here, the first term describes conventional superexchange interaction between the occupied orbitals of Cr atoms mediated by the $sp$ states of iodine ligands. The second term has a more complicated nature and originates from the virtual transitions between the occupied and unoccupied orbitals on two different atoms, which also involves the intraatomic exchange $\cal{J}_H$ (Hund's rule coupling). Schematic representation of the Kugel-Khomskii interaction mechanism is shown in Fig.~\ref{Fig_KugelKhomskii}. In Eq.~(\ref{KKh}), $t_{1}^{t_{2g}-t_{2g}}$ and $t_{1}^{t_{2g}-e_g}$ are effective hoppings between different orbital states of the nearest-neighbor Cr atoms, and $\Delta$ is the crystal-field splitting between corresponding occupied and empty orbitals. All the parameters could in principle be estimated from non-spin-polarized DFT calculations. The Kugel-Khomskii model was successfully applied to reveal the microscopic origin of magnetic interactions and provide its quantitative description in a large variety of systems, including strongly frustrated transition metal oxides \cite{Mazurenko_Na2V3O7}. However, in the case of CrI$_3$ monolayer spin polarization induces strong changes in the hybridization between $d$ states of chromium and $sp$ states of iodine (see Fig.~\ref{Bands_Comparing} for spin up). This results in a significant variation of the hopping integrals, and prevents us from using Eq.~(\ref{KKh}) for a quantitative analysis of the Green's function results.  

On a qualitative level, Eq.~(\ref{KKh}) describes the decay of $t_{2g}-t_{2g}$ exchange interactions for increasing $U$, in agreement with the results shown in Fig.~\ref{Tab_OrbDecomp_Total_per_CS}. On the other hand, nearly constant behavior of the $t_{2g}-e_{g}$ coupling for $U >$ 1 eV can be explained within the Kugel-Khomskii model, if one assumes $U \gtrsim \Delta$. In this case, for not too small $\Delta$ (a direct estimation from Fig.~\ref{DoS_plot} yields $\Delta~\approx$~2~eV), the first term in Eq.~(\ref{KKh}) is not dominant, ensuring an FM state of the system. We emphasize the detailed balance of the two interactions and a crucial role of the Coulomb interaction in determining the ground magnetic state in CrI$_3$. In the situation with $U\gg \Delta$, the second term in Eq.~(\ref{KKh}) would be suppressed, resulting in a net AFM state. We also point to a decisive role of the Hund's rule coupling $\cal{J}_H$ in the formation of CrI$_3$ ferromagnetism. $\cal{J}_H$ is weakly affected by the dielectric screening \cite{Rudenko12}, and is primarily determined by the degree of orbital localization. It is, therefore, not expected that the FM exchange interactions in CrI$_3$ can be further enhanced in the presence of a dielectric environment. We admit, however, that for a deeper microscopic understanding of exchange interactions in this system, more sophisticated many-body approaches may need be involved. Such methods include, for example, dynamical mean field theory and its extensions, allowing for a more accurate and systematic treatment of the electron correlation effects. 

\begin{table}[h!]
\centering
\caption [Bset]{Main contributions to the Cr-Cr exchange interactions (in meV) calculated for two different values of the Coulomb interaction ($U$) in CrI$_3$. 1NN and 2NN denote first and second nearest-neighbor interactions, respectively.}
\label {Jij_OrbDecomp_Table}
\setlength{\tabcolsep}{7pt}
\begin {tabular}{c|c|c|c|c}
 \hline  \hline
 \multirow{2}{*}{$J_{ij}$} & \multicolumn{2}{c}{$U$ = 0 eV} \vline& \multicolumn{2}{c}{$U$ = 3 eV} \\
\cline{2-5}
                           & $t_{2g}-t_{2g}$ & $t_{2g}-e_{g}$ & $t_{2g}-t_{2g}$ & $t_{2g}-e_{g}$ \\
 \hline 
 1NN  &  $-2.743$  &  3.041  &  $-0.569$  &  1.574  \\
 2NN  &  $-0.228$  &  0.857  &  $-0.067$  &  0.460  \\
 \hline  \hline
\end {tabular}
\end {table}

\begin{figure}[h!]
\centering
\includegraphics[width=0.95\textwidth,angle=0]{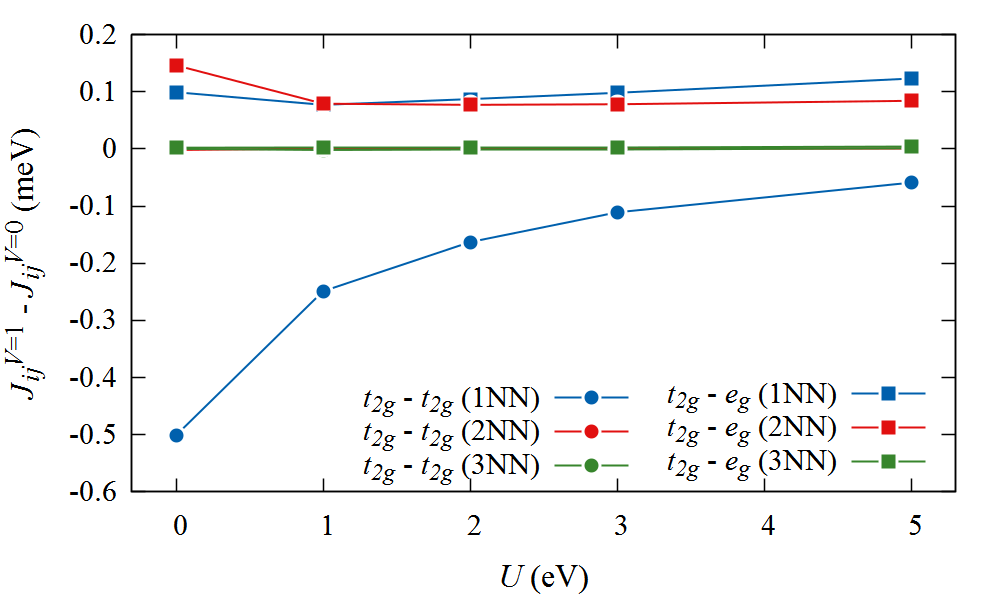}
\caption{The effect of the vertical bias on the Cr--Cr exchange interactions in CrI$_3$, presented as the field-contribution to the exchange $J_{ij}^{V=1}~-~J_{ij}^{V=0}$. Circles and squares denote $t_{2g}-t_{2g}$ and $t_{2g}-e_{g}$ channels, respectively.}
\label{Jij_ESEF_Diff}
\end{figure}

\paragraph{Simulation of electric field.}
Motivated by the ability to control materials' properties by means of the gate voltage in experiments, we examine the influence of external static electric field on monolayer CrI$_3$. To this end, we induce an additional energy gap of 1 eV between the orbitals of I atoms that geometrically reside above and below the Cr atoms, withe the corresponding Hamiltonian given by Eq.~(\ref{Ham_with_VertBias})). At the level of the band structure, the application of the bias voltage results in a slight ($\lesssim 5$\%) narrowing of the band gap. Despite the fact that the net magnetic moment remains almost unchanged, the exchange interactions acquire additional AFM contribution, which is almost independent of the $U$ values (see Fig.~\ref{Fig_JijTotalCS}). 
In Fig. \ref{Jij_ESEF_Diff}, we show an orbitally decomposed difference $J_{ij}^{V=1}~-~J_{ij}^{V=0}$ as a function of $U$, demonstrating overall trend toward destabilization of the FM state in the presence of vertical bias. However, the effect of the two interaction channels is opposite: Unlike the $t_{2g}-t_{2g}$ interaction, the $t_{2g}-e_{g}$ interaction tends to stabilize the FM state. As we well show below, slight modification of exchange interactions in the presence of vertical bias results in the appearance of new excitations in the spin-wave spectrum.


\paragraph{Curie temperature estimation.}
Magnetocrystalline anisotropy is known to be the main factor stabilizing long-range magnetic order in two dimensions. In turn, magnetic anisotropy in CrI$_3$ originates from the strong spin-orbit coupling arising due to heavy iodine atoms \cite{SOC_as_origin_of_Anisotropy}, and can be further stimulated by strain \cite{CrI3_MAE}.  Magnetic anisotropy ensures a gap in the spin-wave spectrum, resulting in a finite Curie temperature, as it immediately follows from Eqs.~(\ref{Formula_Nq}) and (\ref{Formula_Tc_RPA}). To estimate the Curie temperature in CrI$_3$, we consider two magnetic sublattices and
solve the system of equations given by Eq.~\ref{Formula_Tc_RPA}. For the FM ground state, the system has only one solution with $\langle s_{1}^{z} \rangle = \langle s_{2}^{z} \rangle$.

\begin{figure}[h!]
\centering
\includegraphics[width=0.95\textwidth,angle=0]{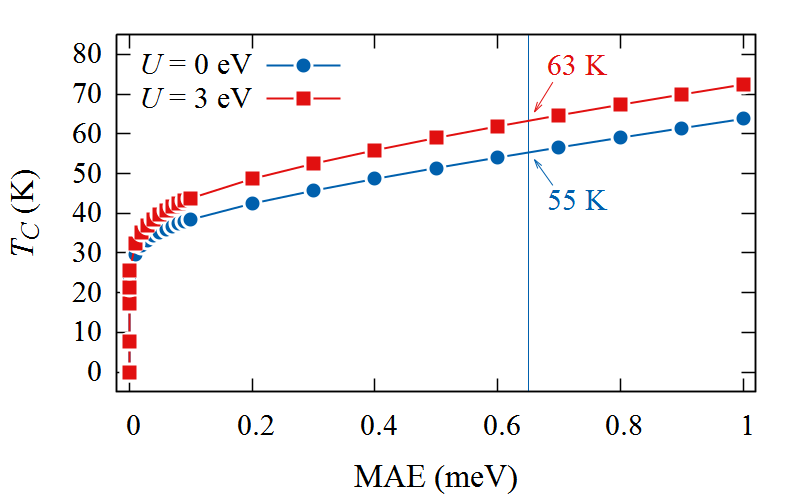}
\caption{The Curie temperature in CrI$_3$ as a function of MAE calculated within RPA for $U$ = 0 eV and $U$ = 3 eV. Blue vertical line denotes MAE~$=0.65$ meV obtained from first-principles calculations in the presence of spin-orbit coupling \cite{Lado_2017}.}
\label{Fig_Tc_RPA}
\end{figure}

As a next step we calculate $T_{C}$ as a function of MAE for $U=0$ and $U=3$ eV, with the results presented in Fig.~\ref{Fig_Tc_RPA}. As expected, for MAE~$\rightarrow 0$ we have $T_{C} \rightarrow 0$ in accordance with the Mermin-Wagner theorem \cite{PhysRevLett.17.1133}, while for MAE~$>0.1$ meV one can see a linear growth of $T_{C}$.
The value of MAE can be estimated from first-principles calculations, which predict MAE~$\approx0.65$ meV \cite{Lado_2017}, shown in Fig.~\ref{Fig_Tc_RPA} by a vertical line. 
The presence of an on-site Coulomb interaction leads to an enhancement of $T_{C}$, as it is expected from the suppression of the AFM $t_{2g}-t_{2g}$ interacting channel. 
Using realistic values for MAE, we obtain $T_{C}=55$~K for \textit{U}~=~0~eV and 63~K for \textit{U}~=~3~eV, which is in reasonable agreement with the experimental extrapolation to the monolayer limit (45~K) \cite{CrI3_Tc_exp}. This allows us to conclude that introducing additional screening of the Coulomb interaction in CrI$_3$ monolayer (e.g., by metallic substrates) is not favorable in terms of magnetic ordering.
It is worth noting that the Curie temperature is found to be almost insensitive to the applied vertical bias:
for non-zero $U$ the difference between $V$ = 0 eV and $V$ = 1 eV does not exceed 1\% for any considered MAE. This observation in consistent with the results of previous theoretical studies \cite{Paper_05}. Nevertheless, we do not exclude further $T_{C}$ stimulation by tuning the balance between $t_{2g}-t_{2g}$ and $t_{2g}-e_{g}$ interaction channels through the crystal field modification induced by strain \cite{CrI3_MAE}, charge doping \cite{PhysRevB.98.155148}, or by formation of heterostructures \cite{doi:10.1021/acs.jpcc.9b04631}.

\begin{figure}[h!]
\centering
\includegraphics[width=0.95\textwidth,angle=0]{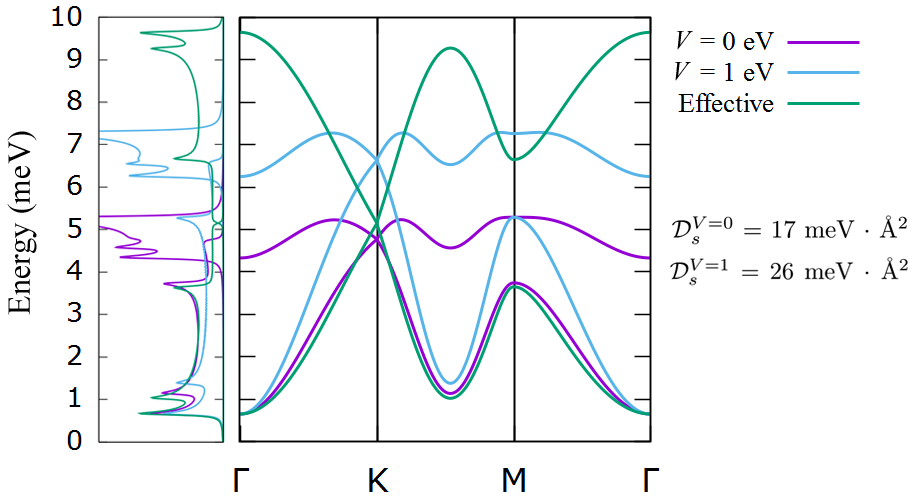}
\caption{Spin-wave dispersion and density of magnon states calculated for CrI$_3$ monolayer using the exchange interaction obtained in this work with $U$ = 3 eV, MAE~$=0.65$ meV, $V$~=~0 and 1~eV, as well as using an effective nearest-neighbor model with $J_{1}^{*} =$ 2.25 meV \cite{SOC_as_origin_of_Anisotropy}. The spin stiffness constant ${\cal{D}}_{s}$, reflecting the curvature near the $\Gamma$ point, is provided for reference.}
\label{Fig_Eq}
\end{figure}

Finally, it is instructive to restore the spin-wave dispersion spectra at zero temperature, and compare it with the effective first nearest-neighbor models proposed earlier. In Fig.~\ref{Fig_Eq}, we show the corresponding spectra $\epsilon({\bf q})$ obtained from the calculated exchange interactions for \textit{U} = 3 eV and MAE~$0.65$ meV with and without the vertical bias, and using the model from Ref.~\onlinecite{SOC_as_origin_of_Anisotropy} with $J_{1}^{*} =$ 2.25 meV. One can see a qualitative difference between the two spectra: the spectrum calculated in our work exhibits narrow bands in the region 5--7 meV depending on the applied bias voltage, which give rise to sharp peaks in the corresponding energy region. The difference between the two spectra points to an important role of exchange interactions at distances beyond the nearest-neighbor, included in our model. 
We also note that the application of vertical electric field modifies the curvature of the spin-wave spectrum near the $\Gamma$ point, i.e. the spin stiffness, being one of the basic characteristics of low-energy magnetic excitations \cite{SpStiffness.Experiment, Kashin.JMMM}. Assuming quadratic form of the spin-wave spectrum at ${\bf q}\rightarrow 0$: $\epsilon({\bf q})\approx \Delta + {\cal D}_S {\bf q}^2$, the spin stiffness constant can be estimated as ${\cal{D}}_{s}^{V = 0}$~=~17 meV~$\cdot$~\AA$^2$ and ${\cal{D}}_{s}^{V = 1}$~=~26 meV~$\cdot$~\AA$^2$, which is in reasonable agreement with previous assessments \cite{PhysRevX.8.041028}. 
One can see that although there is almost no effect of bias voltage on $T_C$, the spin stiffness constant can be increased significantly in the presence of vertical electric field. In our case, for $V=1$ eV it is nearly 50\% higher.
We note, however, that our model does not include the dependence of MAE on the electric field, as well as it does not take into account the field-induced inversion symmetry breaking, which is to be accompanied by the Dzyaloshinskii-Moriya interaction \cite{PhysRevB.97.054416}. These effects lie beyond the scope of our present work, but can be considered in future studies. Our findings suggest a complex character of magnetic excitations in CrI$_3$, which deserve further attention from both experimental and theoretical perspectives.

\section{Conclusion}

We have performed a systematic theoretical study of magnetic properties of monolayer CrI$_3$, material with vast prospects for technological applications in the domain of spintronics.
Starting from first-principles DFT calculations within the GGA+$U$ approach, we constructed a low-energy microscopic model for the electronic structure. This model allowed us to estimate orbital-resolved exchange interactions in CrI$_3$ using the advantage of the magnetic force theorem. 

Orbital decomposition of isotropic exchange interactions revealed an essentially competing character of $t_{2g}-t_{2g}$ and $t_{2g}-e_{g}$ interacting channels. To explain the interaction origin, we made use of the Kugel-Khomskii formalism, which assumes two distinct contributions to the exchange coupling: (i) AFM, governed by the conventional superexchange mechanism involving occupied $t_{2g}$ orbitals, and (ii) FM, arising from the interaction between occupied $t_{2g}$ and unoccupied $e_g$ orbitals, mediated by the intraatomic Hund's rule coupling. 

We also found that the Coulomb interaction is an important factor stabilizing long-range magnetic ordering in CrI$_3$. In the presence of additional dielectric screening, the $t_{2g}-t_{2g}$ FM interacting channel becomes suppressed, leading to a considerable reduction of the Curie temperature. Our estimation within the RPA scheme gives $T_C\approx 63$ K, which is in reasonable agreement with experimental observations. Additionally, we examined the effect of electric field applied normal to the plane of CrI$_3$. Although it is found that the electric field tends to destabilize short-range FM ordering, $T_C$ remains virtually unaffected. 


The calculated exchange interactions allowed us to find interesting details of the spin-wave spectrum. Contrary to the effective nearest-neighbor exchange models, our calculations predict the appearance of intensive spin excitations at energies 5--7 meV, depending on the applied bias voltage. Apart from that, we showed that the spin stiffness constant can be effectively controlled by the external electric field. Summarizing, our findings shed light on the mechanisms behind the magnetic ordering in CrI$_3$, and can motivate further theoretical and experimental studies of magnetic properties of this material and its analogues.
\begin{acknowledgments}
This work was supported by the Russian Science Foundation Grant No. 17-72-20041.
\end{acknowledgments}

\bibliography{mybibfile}

\end{document}